\documentstyle[epsf,epsfig,wrapfig,here,12pt]{article}    % Specifies the document style.
\begin{document}           % End of preamble and beginning of text.
\baselineskip=0.33333in
\begin{quote} \raggedleft TAUP 2831-06
\end{quote}
%\title{A Sample Document}  % Declares the document's title.
%\author{Leslie Lamport}    % Declares the author's name.
%\date{December 12, 1984}   % Deleting this command produces today's date.
\vglue 0.5in
%\maketitle                 % Produces the title.
\begin{center}{\bf Special Relativity is\\
an Excellent Theory}
\end{center}
\begin{center}E. Comay$^*$
\end{center}

\begin{center}
School of Physics and Astronomy \\
Raymond and Beverly Sackler Faculty of Exact Sciences \\
Tel Aviv University \\
Tel Aviv 69978 \\
Israel
\end{center}
\vglue 0.5in
\vglue 0.5in
\noindent
\vglue 0.2in
\noindent
Abstract:

Criteria for defining errors of a physical theory are formulated. It is
shown that the Special Theory of Relativity (STR) has a solid mathematical
basis. An enormous amount of experiments carried out in particle
physics use beams of particles having
a very high energy. The data of these
experiments are consistent with STR and support our confidence that
STR is an excellent theory. Several specific cases of this issue are
discussed explicitly. Contrary to a common belief, it is proved
that the contemporary mainstream of physicists 
adhere to some theoretical ideas that violate STR.

\newpage

{\bf 1. Introduction}
\vglue 0.3333333in

   The validity of physical
theories should be tested time and again. Such a practice
enables the increase of our confidence in good theories and the removal
of erroneous ones. In order to carry out this task, one needs to define
the structure of physical theories and their interrelations.
Criteria for errors in physical theories can be created on this
basis. This work presents the fundamental elements of the Special
Theory of Relativity (STR) and explains why it should be regarded as
a self-consistent and excellent theory. STR is used in classical
physics and in quantum physics as well. The main part of the
discussion carried out in this work is restricted to the validity
domain of classical physics.

The second Section discusses the general structure of physical theories
and defines criteria for a rejection of a theory because of its erroneous
properties. The third Section presents fundamental elements of STR
pertaining to
mechanics and to electrodynamics. The fourth Section examines
some peculiar (and counterintuitive) predictions
of STR and shows that these
predictions are consistent with experimental data. Several examples
proving that some widely accepted contemporary physical theories are
inconsistent with STR, are discussed in the fifth Section. The last
Section contains concluding remarks.

In this work, Greek indices run from 0 to 3 and Latin indices run
from 1 to 3.
Units where $\hbar = c = 1$ are used. In this unit system,
the celebrated relativistic formula $E=mc^2$ reduces to $E=m$. For
these reasons, the symbol $c$ is removed in many cases and
the symbol $m$ denotes not the dynamic mass but the
particle's mass in its instantaneous rest frame.
The relativistic factor $\gamma = (1 - v^2)^{-1/2}$. The symbol
$_{,\mu}$ denotes the partial differentiation with respect to $x^\mu $.

\vglue 0.6666666in
{\bf 2. The Structure of Physical Theories}
\vglue 0.3333333in

   A physical theory resembles a mathematical theory. Both rely on a set
of axioms and employ a deductive procedure for yielding theorems, corollaries,
etc. The set of axioms and their results are regarded as elements of the
structure of the theory.
However, unlike a mathematical theory, a physical
theory is required to explain existing experimental data and to predict results of
new experiments.

   This distinction between a mathematical theory and a physical theory
has several aspects. First, experiments
generally do not yield precise values but
contain estimates of the associated errors. (Some quantum mechanical
data, like spin, are the exception.) It follows that in many cases,
a certain numerical difference between theoretical predictions and
experimental data is quite acceptable.

   Next, one does not expect that a physical theory should explain every
phenomenon. For example, it is well known that physical theories yield
very good predictions for the motion of planets around the sun. On the
other hand, nobody expects that a physical theory
be able to predict the
specific motion of an eagle flying in the sky. This simple example proves
that the validity of a physical theory should be evaluated only with
respect to a limited set of experiments. The set of experiments which
are relevant to a physical theory is called its domain of validity. (A
good discussion of this issue can be found in [1], pp. 1-6.)

   Relations between two physical theories can be deduced from an
examination of their domain of validity. In particular, let
$D_A$ and $D_B$
denote the domains of validity of theories $A$ and $B$,
respectively. Now, if $D_A$ is a subset of $D_B$ then
one finds that the rank of theory
$B$ is higher than that of theory $A$ (see
[1], pp. 3-6). Hence, theory $B$ is
regarded as a theory having a more
profound status. However, theory $A$ is not ``wrong", because it yields
good predictions for experiments belonging to its own (smaller) domain
of validity. Generally, theory
$A$ takes a simpler mathematical form. Hence,
wherever possible, it is used in actual calculations.
Moreover, since theory $A$ is good in its validity domain
$D_A$ and $D_A$ is a part of $D_B$
then one finds that {\em theory A imposes
constraints on theory
B, in spite of the fact that B's rank is higher than
A's rank}. This self-evident
relation between lower rank and
higher rank theories is called here ``restrictions imposed by a lower rank
theory."
Thus, for example, although Newtonian
mechanics is good only for cases where the velocity $v$ satisfies
$v/c \rightarrow 0$, relativistic mechanics should yield formulas
which agree with corresponding formulas of Newtonian mechanics,
provided $v$ is small enough. As is very well known, STR satisfies this
requirement.

   Having these ideas in mind, a theoretical error is regarded here as a
mathematical part of a theory that yields predictions which are clearly
inconsistent with experimental results, where the latter are carried out
within the theory's validity domain. The direct meaning of this definition
is obvious. It has, however, an indirect aspect too. Assume that
a given theory has a certain part, $P$, which is regarded as well
established. Thus, let $Q$ denote
another set of axioms and formulas which hold in 
(at least a part of) $P$'s domain of validity. Now, assume that 
$Q$ yields predictions that are inconsistent with those of $P$ and
the inconsistency holds in the common part of their domains of validity.
In such a case, $Q$ is regarded as a
theoretical error. (Note that, as explained above, $P$ may belong to a lower
rank theory.) An error in the latter sense is analogous to an
error in mathematics, where two elements of a theory are inconsistent
with each other.

   There are other aspects of a physical theory which have
a certain value but are not well defined.
These may be described as neatness, simplicity and physical acceptability of
the theory. A general rule considers theory $C$ as simpler (or neater)
than theory $D$ if theory $C$ relies on a smaller number of axioms.
These properties of a physical theory are relevant to a
theory whose status is still undetermined because there is a lack of
experimental data required for its acceptance or rejection.

   The notions of neatness, simplicity and physical acceptability have
a subjective nature and so it is unclear how disagreements based on them
can be settled. In particular, one should note that ideas concerning
physical acceptability changed dramatically during the 20th century. Thus,
a 19th century physicist would have regarded many well established
elements of contemporary physics as unphysical. An incomplete list of
such elements contains the relativity of length and time intervals, the
non-Euclidean structure of
space-time, the corpuscular-wave nature of pointlike particles,
parity violation and the nonlocal nature of quantum mechanics (which is
manifested by the EPR effect).

   For these reasons neatness, simplicity and physical acceptability
of a theory have a secondary value. Thus, if there is no further evidence,
then these aspects should not be used for taking a {\em final decision}
concerning the
acceptability of a physical theory.

   Before concluding these introductory remarks, it should be stated that the
erroneous nature of a physical theory $E$ cannot be established merely
by showing the
existence of a different (or even a contradictory) theory $F$. This point
is obvious. Indeed, if such a situation exists then one may conclude that
either of the following relations holds: the two theories agree/disagree on
predictions of experimental results belonging to a common
domain of validity. If the theories agree on
all predictions of
experimental results then they are just two different mathematical
formulations of the {\em same} theory. (The Heisenberg and the
Schroedinger pictures of quantum mechanics are an example of
this case.) If the theories disagree then
(at least) theory $E$
{\em or} theory $F$ is wrong. However, assuming that neither $E$ nor
$F$ relies on a mathematical
error, then one cannot decide on the issue without having an adequate
amount of experimental data.

   Another issue is the usage of models and phenomenological formulas.
This approach
is very common in cases where there is no established theory or where
theoretical formulas are
too complicated. A model is evaluated by its usefulness and not by its
theoretical correctness. Hence, models apparently do not belong to
the subject of this compilation of Articles.

\vglue 0.6666666in
{\bf 3. The Mathematical Structure of the Special Theory of Relativity}
\vglue 0.3333333in

Within the scope of this work, one certainly cannot write a
comprehensive presentation of STR. As a matter of fact, there is no need
for doing that, because there are many good textbooks on this subject.
References [2,3] as well as many other textbooks may be used
by readers who are still unacquainted with STR. Hence,
fundamental elements of the mathematical structure of STR are presented
here without a thorough pedagogical explanation.

STR is based on 2 postulates:
\begin{itemize}
\item[{1.}] The laws of mechanics and of electrodynamics take the same
form in all inertial frames.
\item[{2.}] The speed of light in vacuum takes the same value $c$ in
all inertial frames (and it is independent of the velocity of the source).
\end{itemize}

The theory derived from these postulates can be formulated by using
tensor calculus within Minkowski space of 4 dimensions. Three equivalent
forms of this space can be found in the literature.
In these forms the metric tensor (denoted by $g_{\mu \nu }$) is diagonal and
contains the numbers $\pm 1$. The signature of the three forms takes the
values 4, 2 and $-2$, respectively. In the signature 4, the metric is the
unit tensor and calculations use complex numbers. The metric used here
is (1,-1,-1,-1). Apparently, this is the most popular metric used by
modern textbooks.

The differential of the interval $ds$ is obtained from $ds^2 = dt^2 - dx^2$.
Lorentz transformations are second rank tensors $L^\mu _\nu $
that conserve the length of the interval. They are used for transforming
quantities from one inertial frame to another. Lorentz transformations
form a group. A subgroup of this group is the group of rotations in
the ordinary 3-dimensional space. The Poincare group is the group
that contains the Lorentz group and the group of space-time translations.

There are some
important physical quantities which
are invariant under Lorentz transformations
(these invariants are also called Lorentz scalars).
These invariants are the interval;
the following relation of
energy and momentum components of
a closed system $E^2 - P^2$;
$B^2 - E^2$ and ${\bf E\cdot B}$ of the electromagnetic fields.
The electric charge is a Lorentz scalar too.

Some other physical quantities are entries of first rank tensors (also
called 4-vectors). Thus, space-time coordinates are entries of
a 4-vector denoted by $x^\mu $.
For coordinates of the path of a moving massive particle, the
square of the interval $ds^2 =dt^2 - dx^2 > 0$. Hence, the 4-velocity of a
massive particle $v^\mu \equiv dx^\mu/ds=\gamma (1,{\bf v})$
is a well defined 4-vector. Similarly, the 4-acceleration is defined
as follows $a^\mu \equiv dv^\mu/ds$. Energy and momentum of a closed
system are entries of
the 4-vector $P^\mu \equiv (E,{\bf P})$. The scalar and
vector potentials of electrodynamics are entries of the 4-vector
$A^\mu \equiv (\Phi,{\bf A})$. The 4-current is another 4-vector. Here
$j^\mu \equiv (\rho,\rho {\bf v})$, where $\rho $ denotes charge
density. This 4-current satisfies the
continuity equation $j^\mu _{,\mu}=0$, which proves charge conservation.
The 4-current can be written in a different notation, where $\rho $
denotes probability density and all entries
of the 4-vector are multiplied by
the electric charge $e$.
An analogous 4-vector is the mass current where the rest mass $m$
(which is a Lorentz scalar!) replaces the electric charge.

Electromagnetic fields are components of a second rank antisymmetric
tensor which is the 4-curl of $A_\mu $. Thus
$F_{\mu \nu } \equiv A_{\nu ,\mu} - A_{\mu ,\nu }$. Energy and
momentum densities as well as energy and momentum currents
are entries of a second rank symmetric tensor
$T^{\mu \nu }$. This tensor is called the energy-momentum tensor (or
the stress energy tensor).
Thus, $T^{00}$ is the energy density and $T^{i0}$ are
densities of momentum components.

The density of angular momentum components
are entries of a third rank tensor
$S^{\lambda \mu \nu} \equiv x^\lambda T^{\mu \nu} - x^\mu T^{\lambda \nu}$.

It is interesting to note that Maxwellian electrodynamics predicts the
existence of transverse electromagnetic waves that satisfy the following
equation
\begin{equation}
\frac {\partial ^2{\bf E}}{\partial t^2} - \nabla ^2 {\bf E} = 0
\label{eq:MAXWELLWAVES}
\end{equation}
and a similar equation for the components of the magnetic field.
In the vacuum, these waves travel
in the speed of light. Moreover, since
Maxwell's wave equation is independent of quantities of the 
inertial frame
where the fields are measured (and of the velocity of the source 
of the fields as well), one concludes that Maxwellian fields
travel in the speed of light $c$ in all frames. This conclusion
agrees completely with postulate 2 of STR.

The mathematical structure of Minkowski space is known to be
self-consistent. Moreover, as stated above, STR agrees with Newtonian
mechanics in cases where $v/c\rightarrow 0$.
Thus, the mathematical aspect of STR is flawless and its
validity should be examined by means of a
comparison of its predictions with well established experimental data.

\vglue 0.6666666in
{\bf 4. Experimental Data and Special Relativity}
\vglue 0.3333333in

As explained in Section 2, the acceptability of STR should be examined within
its validity domain. Thus experiments where effects of gravitational
field or of noninertial frames can be ignored are examined. Hence,
terrestrial experiments of strong, electromagnetic and weak
interactions belong to the validity domain of STR. This
section discusses several results of STR, some of which may look strange
to everybody who follows his intuition (which has been developed on the
basis of life experience in a macroscopic world and where $v/c \ll 1$).

\begin{itemize}

\item[{1.}] It is proved in STR that the speed of light is an upper
bound for the velocity of massive particles $v<c$. This property is verified
in many experiments. Take for example the CERN's LEP accelerator where
beams of electrons and positrons are accelerated to a very
high kinetic energy. The beams collide and their center
of mass energy exceeds
200 GeV [4]. Thus, electrons and positrons of the beams have
kinetic energy which is more than 200000 times $mc^2$. In spite of this
gigantic kinetic energy, particles do not move faster than light.

Another kind of information are the neutrinos measured from the 1987A
supernova. This supernova exploded about 164000 years ago (data taken
from the Internet site of Wikipedia). Thus the number of seconds
elapsed is about $5\cdot 10^{12}$. On earth,
the neutrino burst lasted about 13
seconds. A variation in the energy of these neutrinos is expected to
hold, due to Doppler shift and other reasons. According to recent
experimental measurements, neutrinos are massive particles
(see [5], pp. 451-467). Therefore,
one may conclude that the variation in speed of these very high energy
particles is less than $10^{-11}$ of their mean speed. This conclusion is
consistent with STR. Indeed, in STR the speed of all very high energy
massive particles is $c(1 - \varepsilon )$, where $\varepsilon $ is a
very small positive number.

\item[{2.}] The equivalence of mass and energy is another result of STR.
This conclusion is seen in many experiments of particle physics.
Thus, the
positronium is a bound state of an electron and a positron. These particles
annihilate each other
and two or three photons are emitted. Photons are massless
particles found in electromagnetic radiation. Hence, they are a form
of energy (which can be converted into heat, etc.). Similarly, the particle
$\pi ^0$ disintegrates into 2 photons. Another experimental example of
the equivalence of mass and energy
is the heat released from a fission of heavy nuclei
like $^{235}U$ and $^{239}Pu$. Here the sum of the masses of the nuclei
produced by fission is smaller than that of the original nucleus.
The difference between the masses appears as a kinetic energy which is
eventually converted into heat.

Processes taking the opposite direction are seen too. Thus, photons
having energy greater than 1 MeV are absorbed by matter in a process called
pair production, where an electron and a positron are created [6]. In
higher energy processes, meson production [7] (namely a $\bar {q}q$
bound state) is observed. In even higher energy, a pair of proton-antiproton
are produced [8].

\item[{3.}] The Lorentz contraction of length is another result of STR. Thus,
a rod of length $l$ looks shorter, if it is measured in an inertial
frame $\Sigma $
where it moves in a direction which is not perpendicular to its
length. Lorentz contraction
is seen in an examination of $\mu $ mesons
having a very high energy. The
half-life time of these particles is about $2.2\cdot 10^{-6}$ seconds.
This time interval should be measured in the particle's rest
frame $\Sigma '$.
Hence, if Lorentz contraction does not hold, then after moving
4000 meters, their number should be about $1.5\%$ of their original number.
After passing 10000 meters, the number should be less than $10^{-4}$
of the original number.
Now, many $\mu $ mesons are produced at the upper part of the atmosphere
as a result of interactions initiated by a very energetic cosmic ray and
a considerable part of these particles
reach sea level. This effect is explained by measuring the time (and
the half-life time) in the particle's rest frame $\Sigma '$ and by
the Lorentz contraction of
the distance between the upper part of the atmosphere and sea level,
which holds in $\Sigma '$.

This effect can also be seen in a $\mu $ meson machine where processes are
under control [9]. Here high energy $\mu $ mesons move in a storage
ring. Lorentz contraction of length in the $\mu $ meson's 
instantaneous rest frame is
seen as a time dilation in the laboratory frame. Thus, in this specific
case, the time dilation factor is about 30. 
This outcome is a very convincing argument supporting the Lorentz contraction
of length.

\item[{4.}] Landau and Lifshitz use STR and prove that
an elementary classical particle must be pointlike (see [2], pp. 43-44). 
This
result is supported by quantum mechanics and by quantum field theory.
Indeed, in these theories
the wave function/field function $\psi (x^\mu)$ depends
on a {\em single} set of space-time coordinates $x^\mu $. Hence, these
functions describe pointlike particles. Experimental results of
the elementary Dirac particles: electrons, $\mu $ mesons and $u,\;d$ quarks are
consistent with this property. This conclusion is inferred from the 
experimental support of the Bjorken scaling in very high energy
scattering [10].

\end{itemize}

   The foregoing examples show several kinds of experimental data,
all of which
are predicted by STR. In addition to these examples, it
can also be stated that an enormous number of experiments
in high energy physics have been carried out during the last 50 years.
These experiments are designed, constructed and analyzed in accordance with
the laws of STR. Therefore, beside yielding specific results, these
experiments provide a solid basis for our confidence that STR is an
excellent theory.

\vglue 0.6666666in
{\bf 5. Violations of the Special Theory of Relativity by Contemporary
Theoretical Ideas}
\vglue 0.3333333in

This Section shows three examples where theoretical ideas
adopted by the mainstream of contemporary physics are inconsistent with STR.

\begin{itemize}

\item[{1.}] The data of high energy photons interacting with nucleons show
that in this case, protons and neutrons are very much alike [7]. These
data cannot be explained by an analysis of the photon interaction with
the electric charge of nucleon constituents.
Thus, an idea called Vector Meson Dominance (VMD) has been suggested for
this purpose.

The main point of VMD is that the wave function of an energetic photon
takes the form
\begin{equation}
\mid \gamma >\; = c_0\mid \gamma _0> + c_h \mid h>
\label{eq:GAMMA}
\end{equation}
where $\mid \gamma >$ denotes the wave function of a physical photon,
$\mid \gamma _0>$
denotes the pure electromagnetic
component of a physical photon and $\mid h>$ denotes
its hypothetical hadronic component. $c_0$ and $c_h$ are appropriate
numerical coefficients 
{\em whose values depend on the photon's energy} [7,11]. Thus, 
for soft
photons $c_h = 0$ whereas it
begins to take a nonvanishing value for photons whose
energy is not much less then the $\rho $ meson's mass.

The fact that the Standard Model has no other explanation for the hard
photon-nucleon interaction is probably the reason for the survival of
VMD. An analysis published
recently proves that VMD is inconsistent with many well established
elements of physical theories [12]. In particular, VMD is inconsistent
with Wigner's analysis of the Poincare group [13,14]. This outcome
proves that VMD violates STR.

%%%%%%%%%%%%%%%%%%%
\begin{figure}[h]
\vspace*{3ex}
\begin{center}
\rotatebox{0} {\includegraphics*[height=5cm]{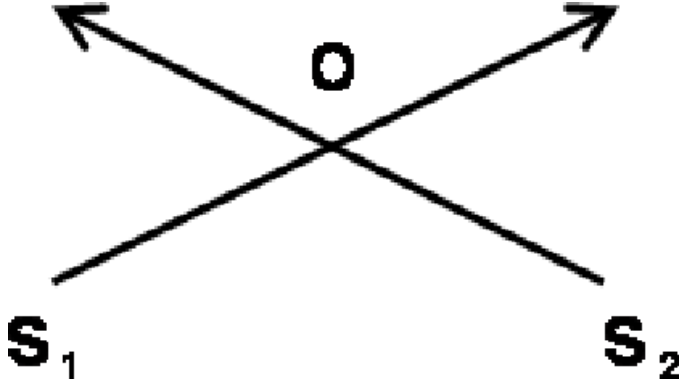}}
% \hspace{10ex}
% \includegraphics*[height=5cm]{rotor.eps}
\end{center}
\begin{quotation}
\caption{
Two rays of light are emitted from sources $S_1$
and $S_2$ which are located at $x=\pm 1$,
respectively. The rays intersect at point $O$
which is embedded in the $(x,y)$ plane.
(This figure is published in [12] and is used here with permission.)
}

% {\it Rotations.} The small circle marks the origin. The point
% $X=(\half,\half,\half,\half)$ is marked by a cross.
%Thick squares show the blocking pattern.

 %\lbl{figprop}}
\end{quotation}
\vspace*{-1ex}
\end{figure}
%%%%%%%%%%%%%%%%%%%

This conclusion can also be proved by the following specific example.
Consider the experiment described in figure 1.
In the laboratory frame $\Sigma $ of fig. 1, the optical
photons of the rays do not interact. Thus, neither energy
nor momentum are exchanged between the rays. Therefore,
after passing through $O$, the photons travel in their
original direction. Let us examine the situation in a frame $\Sigma '$. In
$\Sigma $, frame $\Sigma '$ is seen moving
very fast in the negative direction of
the Y axis. Thus, in $\Sigma '$, photons of the two rays are very
energetic. Hence, if VMD holds then photons of both rays contain hadrons
and should exchange energy and momentum at point $O$.
This is a contradiction because if the
rays do not exchange energy and momentum in frame $\Sigma $ then they
obviously do not do that in any other frame of reference. Thus, this
simple example proves that VMD violates STR.

\item[{2.}] The Yukawa interaction is derived from the interaction
term of a Dirac spinor $\psi (x^\mu )$
with a Klein-Gordon (KG) particle $\phi (x^\mu )$
(see [15], p.79 and [16], p. 135)
\begin{equation}
L_{Yukawa} = L_{Dirac} + L_{KG} - g\bar {\psi}\psi \phi .
\label{eq:LYUKAWA}
\end{equation}
Here the KG particle plays a role which is analogous to that of the
photon in electrodynamics. The following argument proves that a Lorentz
scalar (like the KG particle) cannot be used as a basis for a field of
force.

Consider the following Lorentz scalar $v^\mu v_\mu$. As a scalar, it
takes a fixed value in all inertial frames. (In the units used here
its value is unity.) Differentiating this expression with respect
to the interval, one finds
\begin{equation}
\frac {d(v^\mu v_\mu )}{ds} = 2v^\mu a_\mu = 0.
\label{eq:VA}
\end{equation}
This relation means that in STR
the 4-velocity is orthogonal to the 4-acceleration.

   Let an elementary classical particle $W$
move in a field of force. The field quantities are independent of the
4-velocity of
$W$ but the associated 4-force must be orthogonal to it. In
electrodynamics this goal
is attained by means of the Lorentz force.
In this case, one finds
\begin{equation}
a^\mu v_\mu = \frac {e}{m}F^{\mu \nu}v_\nu v_\mu = 0,
\label{eq:LOROK}
\end{equation}
where the null result is obtained from the antisymmetry of
$F^{\mu \nu }$ and the
symmetry of the product $v_\mu v_\nu $. In electrodynamics, the
antisymmetric field tensor
$F^{\mu \nu }$ is constructed as the 4-curl of the 4-potential $A_\mu $. Such a
field of force cannot be obtained from the {\em scalar} KG field. 
Now, the notion of force holds in classical physics. Hence, the
classical limit of the Yukawa interaction is inconsistent with STR.

\item[{3.}] Following historical ideas, $\pi $ mesons are regarded as KG
particles (see [15], pp. 79, 122).
This is certainly wrong because it has recently been proved
that the KG equation is inconsistent with well established physical
theories [17,18]. This conclusion is in accordance with Dirac's
negative opinion on the KG equation [19,20].

This matter has also an indirect
aspect pertaining to STR. Indeed, as shown in
point 4 of Section 4, STR proves that
a truly elementary classical particle should be pointlike.
This result is also obtained from the quantum mechanical wave
function $\Psi(x^\mu )$ which depends on a {\em single} set of
space-time coordinates.
Now, the KG equation, is supposed to be a quantum mechanical equation.
As such, it must describe pointlike particles. On the other hand, it
is now recognized that $\pi $ mesons are not pointlike and that their
size is not much smaller than the size of the proton (see [5],
pp. 499, 854.). Therefore
the usage of $\pi $ mesons as KG particles violates STR indirectly.

\end{itemize}

\vglue 0.6666666in
{\bf 6. Concluding Remarks}
\vglue 0.3333333in

The notion of a theoretical error is defined. It is explained that STR
has a solid mathematical basis. The fact that
its formulas agree with Newtonian
mechanics in cases where $v/c \rightarrow 0$ proves that it satisfies restrictions
imposed by a lower rank theory.
Next, it is shown that some peculiar predictions
of STR are confirmed by experiments. The predictions discussed here
are the relation $v<c$ where $v$ denotes the velocity of a massive
particle; the equivalence of mass and energy; the Lorentz contraction;
and the pointlike nature of elementary particles.
The enormous number of experiments
carried out in particle physics use particles whose velocity is in
the relativistic domain where $0 < 1 - v/c \ll 1$. The design, construction
and analysis of these experiments abide by the laws of STR.
The data obtained are compatible with STR and
provide a solid basis for our confidence that STR is an excellent
theory. 

The discussion carried out above concetrates on phenomena belonging to
classical physics. It should be noted that the Dirac equation is
a relativistic quantum mechanical equation.
It predicts correctly the spin of
the electron and the existence of antiparticles. It yields very good
predictions for the energy levels of the hydrogen atom and 
for the electron's g-factor. Corrections to these values
are obtained from quantum field theory, which is a higher
relativistic theory.

It is also proved that, contrary to a common belief, some
theoretical ideas, adopted by the
mainstream of contemporary physicists, violate STR. These ideas are
VMD, the Yukawa theory of a field of force carried by a scalar meson
and the idea that $\pi $ mesons are Klein-Gordon particles.

%           ?????????????

\newpage
References:
\begin{itemize}

\item[{*}] Email: elic@tauphy.tau.ac.il \\
           Internet site: http://www-nuclear.tau.ac.il/$\sim $elic
\item[{[1]}] F. Rohrlich,  {\em Classical Charged Particles}, (Addison-wesley,
Reading Mass, 1965).
\item[{[2]}] L. D. Landau and E. M. Lifshitz, {\em The Classical
Theory of Fields} (Pergamon, Oxford, 1975).
\item[{[3]}] J. D. Jackson, {\em Classical Electrodynamics} (John Wiley,
New York,1975).
\item[{[4]}] K. Hubner, Phys. Rep., {\bf 403}, 177 (2004).
\item[{[5]}] S. Eidelman et al. (Particle Data Group), Phys. Lett. 
{\bf B592}, 1 (2004). 
\item[{[6]}] K. S. Krane, {\em Introductory Nuclear Physics} (John Wiley, 
New York, 1988). p. 201.
\item[{[7]}] T. H. Bauer, R. D. Spital, D. R. Yennie and F. M. Pipkin,
Rev. Mod. Phys. {\bf 50}, 261 (1978).
\item[{[8]}] P. Achard et al., Phys. Lett., {\bf B571}, 11 (2003).
\item[{[9]}] R. M. Carey et al,, Phys. Rev. Lett., {\bf 82}, 1632 (1999)
\item[{[10]}] D. H. Perkins, {\em Introduction to High Energy Physics}
(Addison-Wesley, Menlo Park, CA, 1987). Pp 272-273.
\item[{[11]}] H. Frauenfelder and E. M. Henley, {\em Subatomic Physics},
(Prentice Hall, Englewood Cliffs, 1991). pp. 296-304.
\item[{[12]}] E. Comay, Apeiron {\bf 10}, no. 2, 87 (2003).
\item[{[13]}] E. P. Wigner, Annals of Math., {\bf 40}, 149 (1939).
\item[{[14]}] S. S. Schweber, {\em An Introduction to Relativistic
Quantum Field Theory}, (Harper \& Row, New York, 1964). pp. 44-53.
\item[{[15]}] M. E. Peskin and D. V.
Schroeder, {\em An Introduction to  Quantum Field Theory} (Addison-Wesley,
Reading, Mass., 1995).
\item[{[16]}] G. Sterman {\em An Introduction to Quantum
Field Theory} (University Press, Cambridge, 1993).
\item[{[17]}] E. Comay, Apeiron, {\bf 11}, No. 3, 1 (2004).
\item[{[18]}] E. Comay, Apeiron {\bf 12}, no. 1, 27 (2005).
\item[{[19]}] P. A. M. Dirac, {\em Mathematical Foundations of Quantum
Theory} Ed. A. R. Marlow (Academic, New York, 1978). (See pp. 3,4).
\item[{[20]}] S. Weinberg  {\em The Quantum Theory of Fields}
(Cambridge, University Press, 1995). Vol. 1, pp. 3-8

\end{itemize}

\end{document}